# Intrinsic diamagnetism in the Weyl semimetal TaAs

Yu Liu,[1,*] Zhilin Li,[2,*] Liwei Guo,[2] Xiaolong Chen,[2,§] Ye Yuan,[1,3] Fang Liu,[1,3] Slawomir Prucnal,[1] Manfred Helm,[1,3] Shengqiang Zhou[1]

1. Helmholtz-Zentrum Dresden-Rossendorf, Institute of Ion Beam Physics and Materials Research, Bautzner Landstraße 400, 01328 Dresden, Germany
2. Research & Development Center for Functional Crystals, Beijing National Laboratory for Condensed Matter Physics, Institute of Physics, Chinese Academy of Sciences, Beijing 100190, China
3. Technische Universität Dresden, 01062 Dresden, Germany

We investigate the magnetic properties of TaAs, a prototype Weyl semimetal. TaAs crystals show diamagnetism with magnetic susceptibility of about $-7 \times 10^{-7}$ emu/(g·Oe) at 5 K. A general feature is the appearance of a minimum at around 185 K in magnetization measurements as a function of temperature, which resembles that of graphite. No phase transition is observed in the temperature range between 5 K and 400 K.

Although particle physics and condensed-matter physics sometimes borrow concepts from each other, it is still surprising that high-energy particles can be revealed in condensed matter with low excited energies. In the beginning, Dirac fermions were found in graphene [1] and then in topological insulators [2, 3]. When Dirac semimetals with fourfold degenerate Dirac nodes on the Fermi surface were predicted [4, 5] and realized [6-8], these quasi-particles appeared in three-dimensional space. Furthermore, if a Dirac node can be divided into two overlapping nodes with opposite chirality, Weyl fermions [9] are expected in such materials, which is coined as Weyl semimetals (WSMs). Recently, the transition-metal mono-phosphides and -arsenides TaAs, TaP, NbAs, and NbP have been predicted to be WSMs [10]. With their successful growth of their single crystals, accumulated evidence has supported this prediction. Both the Weyl nodes and the Fermi arcs have been observed in TaAs [11-13] and NbAs [14]. The spin texture of surface Fermi arcs is further confirmed in TaAs [15]. Chiral-anomaly-induced negative magnetoresistance has been measured in NbP [16], TaAs [17], and TaP [18]. Mobility has been found as high as $5 \times 10^6$ cm$^2$ V$^{-1}$ s$^{-1}$ at 1.85 K in NbP [16] and $1.8 \times 10^5$ cm$^2$ V$^{-1}$ s$^{-1}$ at 10 K in TaAs [17]. One of the merits in the family of transition-metal mono-phosphides and -arsenides compared to other WSMs ($Rn_2Ir_2O_7$ pyrochlore [19], $HgCr_2Se_4$ [20] or multilayers of normal insulators and magnetically doped topological insulators [21]) is that they are nonmagnetic, without the interference of magnetic domains or the spin exchange field so that the Berry curvature remains intact. Yet, although being a key property of these materials, their intrinsic diamagnetism has barely gained any attention. Does the Weyl nodes affect the magnetism in a similar manner as the Dirac points in graphene [22]? Further, as defects cannot be avoided due to the composition deviation during material synthesis, the possibility of defect-induced magnetism should be considered [23]. It is necessary to confirm the absence of magnetic ordering. Besides, magnetic measurements may be considered as a rapid method to pre-select perfect crystals for further investigation.

Here, we present an investigation of the magnetic properties of TaAs crystals. Both TaAs single crystals and polycrystals show diamagnetism with a similar magnetic susceptibility. The general feature is the appearance of a minimum in magnetization measurements as a function of temperature. No phase transition is observed at the temperature between 5 K and 400 K.

TaAs crystals were grown using the chemical vapor transport method. The pre-reacted TaAs polycrystalline precursor mixed with the transporting medium iodine was loaded into a quartz tube. After evacuation and sealing, the samples were kept at the growth temperature around 1300 K for three weeks. The dimensions of single crystals can be up to 4 mm. After basic cleaning, the crystals were confirmed to have NbAs-type body-centered-tetragonal structure by X-ray diffraction. No iodine doping or any second phase is detected within the sensitivity limit. (More details are revealed in another unpublished report [24]. And it is pointed that a lot of defects may exist in as-grown TaAs [25].) Three TaAs single crystals (S1, S2 and S3) and two pieces of polycrystalline TaAs (S4 and S5) are selected for this study. These samples are from several batches as they are relatively big sized so that only one or two pieces are qualified in a single batch. Figure 1 shows the typical optical morphology of one of these samples (S3).

Room temperature Raman spectroscopy with different incident and scattered light polarizations is performed using a laser with wavelength of 532 nm. Fig. 2 shows typical Raman spectra of TaAs crystals from S2 with light polarization directions xx and x'x'. The configuration is the same as that in a previous report [26]. The peak around 252.2 cm$^{-1}$ is consistent with the theoretical result of the A1 optical mode (243.2 cm$^{-1}$) and the previously-observed one (251.9 cm$^{-1}$) [26]. The other peak at 171.0 cm$^{-1}$ is assigned to the mode $B_1(1)$. Only the TaAs structure is identified and no other phase is detected within the sensitivity limit.

The magnetization measurements were performed using a superconducting quantum interference device vibrating sample magnetometer with a sensitivity of 10$^{-7}$ emu. The variations of magnetization with the field (M-H) in the range of -50 kOe < H < 50 kOe at 5 K and 300 K demonstrate only diamagnetism existing in all samples. A typical profile from S1 is shown in Fig. 3(a) and 3(b). The magnetic susceptibility is -

7.51 × 10$^{-7}$ emu/(g·Oe) at 5 K and -8.19 × 10$^{-7}$ emu/(g·Oe) at 300 K, respectively. Figure 3(c) shows the typical temperature dependence of the magnetization (M-T) of S5 at 100 Oe. The zero field cooled (ZFC) and field cooled (FC) M-T coincide without any gap indicating no ferromagnetism or antiferromagnetism in TaAs. The smooth curves further confirm the absence of phase separation or any phase transition up to 400 K. Upon rising temperature the magnetization decreases and then increases above 185 K. This tendency is similar in M-T at 50 kOe shown in Fig. 3(d). The absence of any rapid increase of magnetization upon lowering the temperature rules out the possibility of paramagnetism. The fact that the magnetization minimum stays at the same position suggests that the signal originates from intrinsic diamagnetism. The variation of in the magnetization from 5 K to 185 K (the minimum) is 16%. The diamagnetism and the variation in M-T seem to be different from those in graphene [22] but similar to graphite [27]. The π Berry phase and the linear dispersion in graphene lead to the nonlinear dependence of diamagnetism on field and temperature [22]. However, the periodic arrangement of graphene (graphite) modifies its original properties. In graphite, the nonlinear dependence of diamagnetism on field becomes linear and the dependence on temperature shows a minimum at about 20 K [27]. This could also happen to bulk TaAs, which reveals a minimum in the temperature dependence of magnetization. Effects such as band-to-band transitions can be induced when the dimensionality increases from two to three. Therefore, the magnetic properties of bulk TaAs cannot be understood directly via the Weyl nodes. The method of Luttinger and Kohn could be useful in the understanding of this diamagnetism [28] as it takes into account the effects of band-to-band transitions which play an important role in determining the properties of graphite. Further, the nonlinear dependence on field vanishes implying a low mobility in samples, which will conceal the intrinsic properties of the Weyl nodes. According to the comparison between graphene/graphite and TaAs, to avoid the additional bulk effects and then study the Weyl nodes, it is suggested the samples should be prepared as thin as possible. During the measurements, the samples were washed to remove phenolic resin by isopropanol. The magnetization after washing remains the same, which means that TaAs is stable in isopropanol.

Furthermore, the variations from one sample to another are shown in Fig. 4: a) The variation of magnetic susceptibility at 5 K, b) the variation of magnetic susceptibility at 300 K, c) the variation of the temperature of the magnetization minimum and d) the variation of the magnetization difference between 5 K and around 185 K (the minimum of the M-T curves). Note that sample S2 exhibits somewhat different behavior compared to the other samples. Here, we show its result but we did not count it when we calculated the average value. The deviation of S2 from the others could be due to a slight change in crystallinity which occasionally happens during the growth. The self-doping tunes the carrier concentration and moves the Fermi level, which could affect its magnetism. This also reminds that some undetected doping may drive the samples away from the intrinsic properties of bulk TaAs. It could be an interesting issue worth further exploring.

In conclusion, we have studied three TaAs single crystals and two pieces of polycrystalline TaAs. Magnetization measurements confirm the absence of ferromagnetism, antiferromagnetism or paramagnetism. TaAs crystals show intrinsic diamagnetism with magnetic susceptibility of about $-7 \times 10^{-7}$ emu/(g·Oe) at 5 K. No phase separation or phase transition is detected within the sensitivity limit in the measured temperature range from 5 K to 400 K. The general feature is the appearance of a minimum at around 185 K in the M-T curves. Both diamagnetism and the minimum in the M-T curve can be understood by comparison with graphene and graphite.

The work is financially supported by the Helmholtz Postdoc Programme (Initiative and Networking Fund, PD-146). L. W. Guo and X. L. Chen also thank the support by the Ministry of Science and Technology of China (Grants No. 2011CB932700), and the National Natural Science Foundation of China (Grants Nos.51532010, 91422303, 51202286 and 51472265), the Strategic Priority Research Program (B) of the Chinese Academy of Sciences (Grant No. XDB07020100).

* These authors contributed equally to this work.

§ E-mail: chenx29@iphy.ac.cn

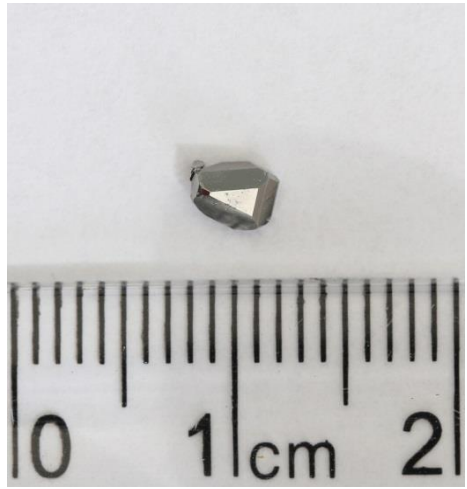

FIG. 1. The optical morphology of sample S3. The ruler is shown for scale.

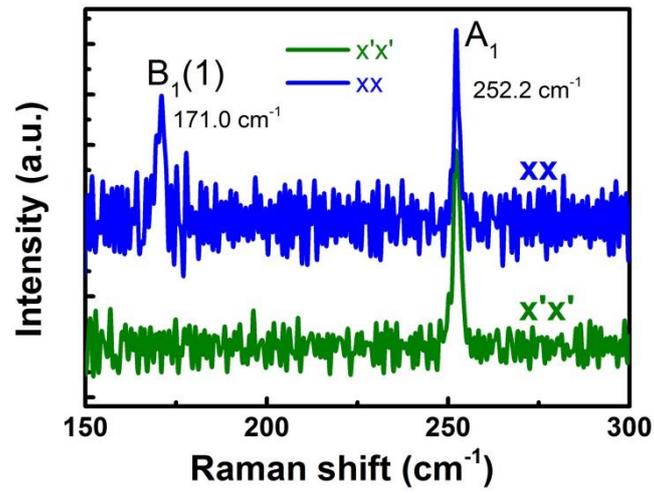

FIG. 2. Room temperature Raman spectra of sample S2 with two incident and scattered light polarizations (xx and x'x').

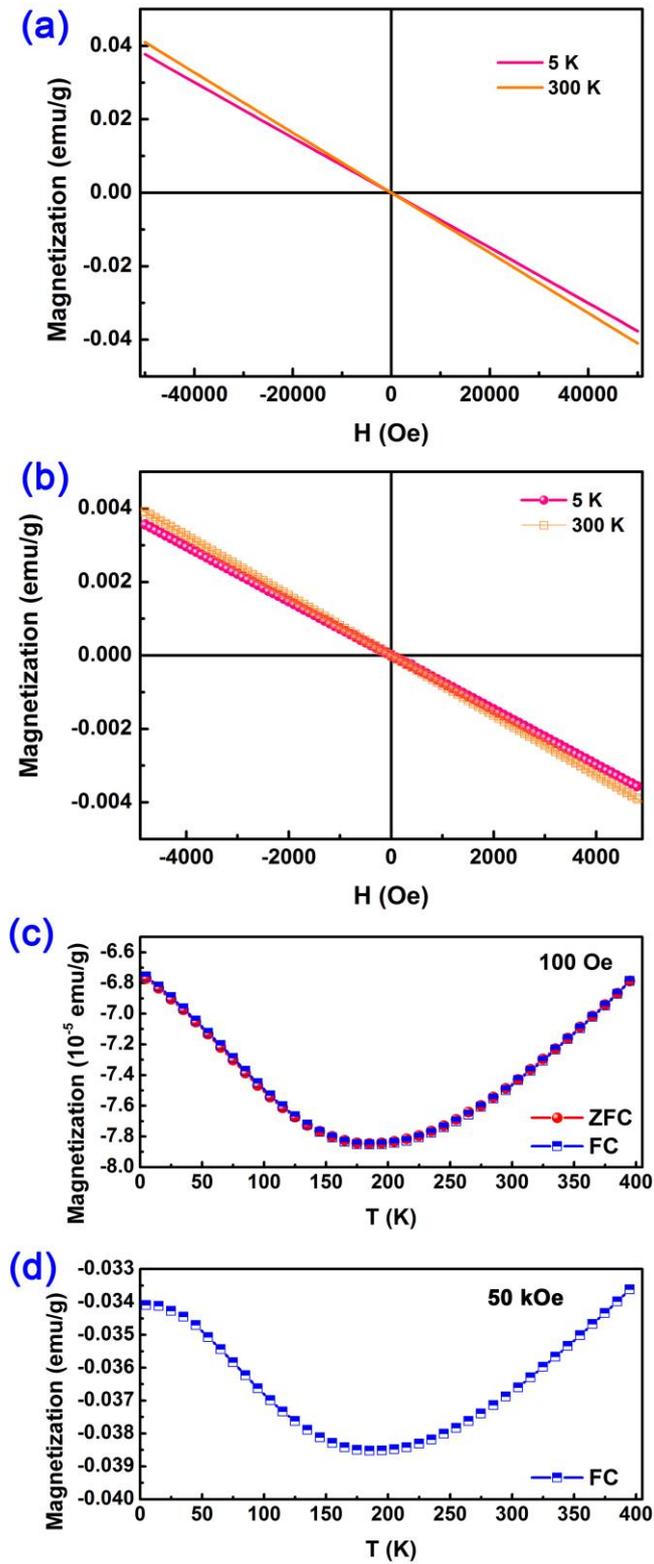

FIG. 3. (a) (b) The magnetization (in units of 1 emu = $10^{-3}$ Am$^2$) measured at T = 5 K or 300 K as a function of magnetic field (1 Oe = $10^3/4\pi$ Am$^{-1}$) in the range -50 kOe < H < 50 kOe or -5 kOe < H < 5 kOe in S1. (c) (d) ZFC / FC magnetization under a field of 100 Oe or FC of 50 kOe varying with temperature from 5 to 400 K in S5.

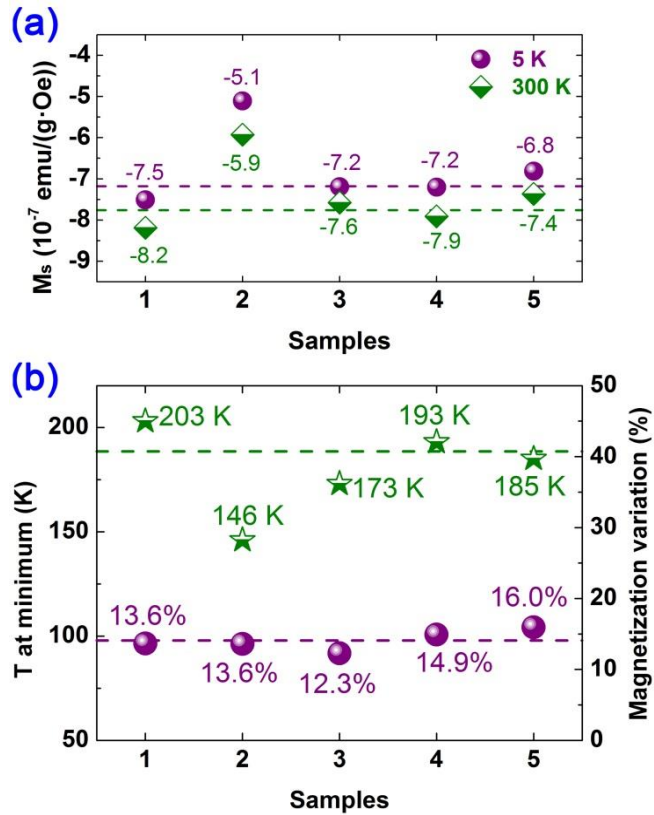

FIG. 4. (a) The variation of magnetic susceptibility at 5 K and the variation of magnetic susceptibility at 300 K. (b) The variation of the temperature of the magnetization minimum and the variation of the magnetization difference between 5 K and around 185 K (the minimum of the M-T curves). The dotted lines are guide to eyes showing the average values. Sample S2 has been excluded in calculating the averages.